\documentclass[twocolumn,aps,superscriptaddress,nofootinbib]{revtex4-1}
\usepackage[dvips]{graphicx}
\usepackage{latexsym,amssymb,amsmath}
\usepackage{color}
\usepackage{enumerate}
\usepackage[bookmarksnumbered,bookmarksopen,colorlinks,citecolor=red,linkcolor=blue]{hyperref}
\usepackage{mathrsfs}
\usepackage{multirow}
\usepackage{bm}
\usepackage[dvipsnames]{xcolor}

\newcommand{\fm}{\textrm{fm}}

\graphicspath{{./figs/}}
\begin{document}   

\title{Detecting nuclear mass distribution in isobar collisions via charmonium}
\author{Jiaxing Zhao}
\email{jzhao@subatech.in2p3.fr}
\affiliation{Physics Department, Tsinghua University, Beijing 100084, China}
\affiliation{SUBATECH, Universit\'e de Nantes, IMT Atlantique, IN2P3/CNRS, 4 rue Alfred Kastler, 44307 Nantes cedex 3, France}

\author{Shuzhe Shi}
\email{shuzhe.shi@stonybrook.edu}
\affiliation{Physics Department, Tsinghua University, Beijing 100084, China}
\affiliation{Center for Nuclear Theory, Department of Physics and Astronomy, Stony Brook University, Stony Brook, New York 11794–3800, USA}

\begin{abstract}
The collective properties of final state hadrons produced in the high statistics $_{44}^{96}$Ru+$_{44}^{96}$Ru and $_{40}^{96}$Zr+$_{40}^{96}$Zr collisions at $\sqrt{s_\mathrm{NN}} = 200~\mathrm{GeV}$ are found to be significantly different. Such differences were argued to be precise probes of the difference in nucleon distribution in the isobar nuclei. 
We investigate the $J/\psi$ production in the isobar collision via a relativistic transport approach. By comparing the isobar systems according to equal centrality bin and equal multiplicity bin, we find that the yield ratio of $J/\psi$ is sensitive to the differences in both the number of binary collisions and the medium evolution.
Besides, the elliptic flow $v_2$ of $J/\psi$ is qualitatively different from the light hadrons, and the ratio between Ru+Ru and Zr+Zr collisions is sensitive to the medium evolution. 
The charmonium production provides an independent probe to study the nucleon distribution in the isobar system.
\end{abstract}
\maketitle

\emph{Introduction.}---
At the mean-field level, nuclear properties like spin, parity of ground state, magic number, and $\beta$ decays can be well described by the nuclear shell model~\cite{Caurier:2004gf}. In the shell model, we know most of the nuclei in their ground state are deformed and non-uniform, except the Doubly-Magic Nucleus. Under the polarization effect of valence nucleons, nuclei with an unfilled shell or sub-shell will be deformed and generates collective motion induced by the interaction between valence nucleons and shell structure~\cite{David:2010}. The deformation depends closely on the number of protons and neutrons in the nucleus and has been studied in low-energy nuclear physics for many years, e.g., see reviews~\cite{RevModPhys.83.1467,WOOD1992101}.  
In relativistic heavy-ion collisions, the nuclei are fully destroyed at the beginning of the collisions, but the deformed information will still show its fingerprint in the final observables~\cite{Rosenhauer:1986tn,Shuryak:1999by}.
The deformed nuclear-nuclear collisions will lead to an anisotropic overlap region. The spatial anisotropy of the collision overlap region in the initial state will transform into a momentum anisotropy of the produced hadrons in the final state~\cite{Ollitrault:1992bk,Voloshin:1994mz,Qiu:2011iv,Filip:2009zz,Alver:2010gr,Carzon:2020xwp}. Therefore, relativistic heavy-ion collisions may supply a new way to probe nuclear deformation.

Recently, a high statistics heavy-ion collision, which collides $_{44}^{96}$Ru+$_{44}^{96}$Ru and $_{40}^{96}$Zr+$_{40}^{96}$Zr with beam energy $\sqrt{s_\text{NN}} = 200~\text{GeV}$, is performed by the STAR Collaboration at the Relativistic Heavy Ion Collider (RHIC)~\cite{STAR:2021mii}\footnote{For the rest of the paper, we will refer to the $_{44}^{96}$Ru and $_{40}^{96}$Zr nucleus as Ru and Zr, separately.}. Such a contrast experiment originally proposed to search for the chiral magnetic effect (CME)~\cite{Kharzeev:2004ey,Kharzeev:2007jp,Fukushima:2008xe}, under the presumption that the same baryon number would lead to the same non-CME background while the different electric number would induce a sizable difference in the CME signal. 
Unexpectedly, the bulk properties, namely the charge multiplicity and elliptic and triangular flows, are found to be different in the isobar systems~\cite{STAR:2021mii}.
While such a subtlety prevents one to make a conclusive statement on the existence of CME in heavy-ion collisions and calls for more efforts in better quantification of the background~\cite{Feng:2021pgf,Kharzeev:2022hqz}, it brings a new opportunity to study the nucleon distribution in relativistic heavy-ion collisions~\cite{Giacalone:2019pca,Giacalone:2020awm,Jia:2021oyt,Jia:2021tzt,Zhang:2021kxj,Jia:2022iji,Giacalone:2021udy,Li:2019kkh,Xu:2021uar,Xu:2021qjw,Xu:2021vpn,Zhao:2022uhl,Bally:2022vgo,Nijs:2021kvn}. 
In such experiments, a new phase of matter, called Quark Gluon Plasma (QGP), is created, and the final state particles can be well explained by a hydrodynamic description of the system evolution, (see e.g. Refs.~\cite{Gale:2013da,Schenke:2010rr,Shen:2014vra}).
The initial state of the evolution is the consequence of multiple nucleon-nucleon collisions, and it provides an unique opportunity of measuring multi-nucleon correlation within a nucleus, regardless of the fact that the mapping between initial condition and final observables is complicated.

\begin{figure}[!htb]
\includegraphics[width=0.4\textwidth]{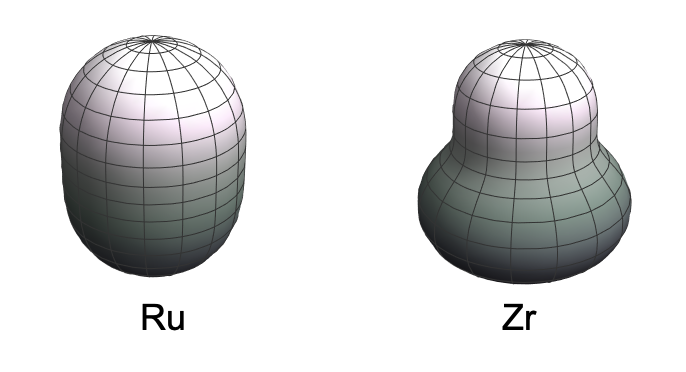}
\caption{3D shapes of Ru and Zr nucleus.}
\label{fig1}
\end{figure}

In Ru and Zr nuclei, the spatial distribution of nucleons are usually described by a deformed Woods--Saxon distribution~\cite{Fricke:1995zz},
\begin{eqnarray}
\rho(r,\theta,\phi)&=&{\rho_0 \over 1+e^{[r-R(\theta,\phi)]/a_0} },\nonumber\\
R(\theta,\phi)&=&R_0(1 + \beta_2 Y_{2,0} + \beta_3  Y_{3,0} +...),
\label{eq.ws}
\end{eqnarray}
where the parameters $R_0$ and $a_0$ are respectively called radius and diffusiveness. The maximum nuclear density $\rho_0$ is given by the normalization condition that $\int \rho(r,\theta,\phi) \mathrm{d}^3\boldsymbol{r} = A$. $Y_{\ell,m}$'s are spherical harmonics.  $\beta_2$ and $\beta_3$ control the quadrupole and octupole deformations, respectively. Both lower-energy nuclear experiments and high-energy heavy-ion collisions suggest that the Ru nucleus has a sizable quadrupole deformation while the Zr nucleus has an octupole deformation ~\cite{Mach:1990zz,HOFER1993173,Zhang:2021kxj,Xu:2021uar}.
Particularly, it is found in~\cite{Zhang:2021kxj} and~\cite{Nijs:2021kvn}, which respectively describes the medium evolution by a particle transport model and hydrodynamics, that the high-precision measurements in the isobar collisions can be described by the parameter set $R_0^\mathrm{Ru}=5.09~\fm$, $a_0^\mathrm{Ru}=0.46~\fm$, $\beta_2^\mathrm{Ru}=0.162$, $\beta_3^\mathrm{Ru}=0.0$, 
$R_0^\mathrm{Zr}=5.02~\fm$, $a_0^\mathrm{Zr}=0.52~\fm$,  $\beta_2^\mathrm{Zr}=0.06$, and $\beta_3^\mathrm{Zr}=0.2$. The corresponding three-dimensional shapes are shown in Fig.~\ref{fig1}.

Charmonium states like $J/\psi$, which are boundstates of a charm quark and its anti-quark, are independent probes of the initial condition. 
Owing to the heavy mass, charm quarks and charmonium states are dominantly produced in the initial hard scattering in heavy-ion collisions. 
After the initial production, they interact with the QGP and their properties get modified. 
The spectrum and flow information of charmonium states in heavy-ion collisions reflect both the QGP properties and initial condition~\cite{Matsui:1986dk,Grandchamp:2002wp,Zhao:2010nk,Du:2015wha,He:2021zej,Yan:2006ve,Liu:2009nb,Zhou:2014kka,Chen:2018kfo,Zhao:2021voa,Chen:2019qzx,Blaizot:2015hya,Katz:2015qja}.
In heavy-ion collisions, the initially produced charmonium bound states might dissociate into scattering states due to the static color-screening effect and dynamic dissociation~\cite{Matsui:1986dk,Satz:2005hx,Peskin:1979va,Bhanot:1979vb,Brambilla:2013dpa}.
Some of them can survive to the end of the QGP evolution and be detected. Another source of final state charmonium particle comes from the recombination of uncorrelated charm and anti-charm quarks in the QGP~\cite{Grandchamp:2002wp,Zhao:2010nk,Du:2015wha,He:2021zej,Yan:2006ve,Liu:2009nb,Zhou:2014kka,Chen:2018kfo}.
We note that initial production rates of both charm quarks and charmonium states are proportional to the number of binary nucleon-nucleon collisions, denoted as $N_\mathrm{coll}$. In the hypothetical situation that the bulk background being the same in the isobar system, the survived charmonium production rate is proportional to $N_\mathrm{coll}$, whereas the recombination production rate is proportional to the square of charm quark number, hence $\propto N_\mathrm{coll}^2$. Compared to light flavor observables, charmonium states are sensitive to different aspects of the initial state and bulk evolution, and henceforth serve as an independent probe of the QGP. 
In this work, we study the differences between the properties of the $J/\psi$ particles produced in Ru+Ru and those in Zr+Zr collisions. We aim to provide an independent probe for the deformations in Ru and Zr nuclei.



\begin{figure*}[!htb]
\includegraphics[width=0.4\textwidth]{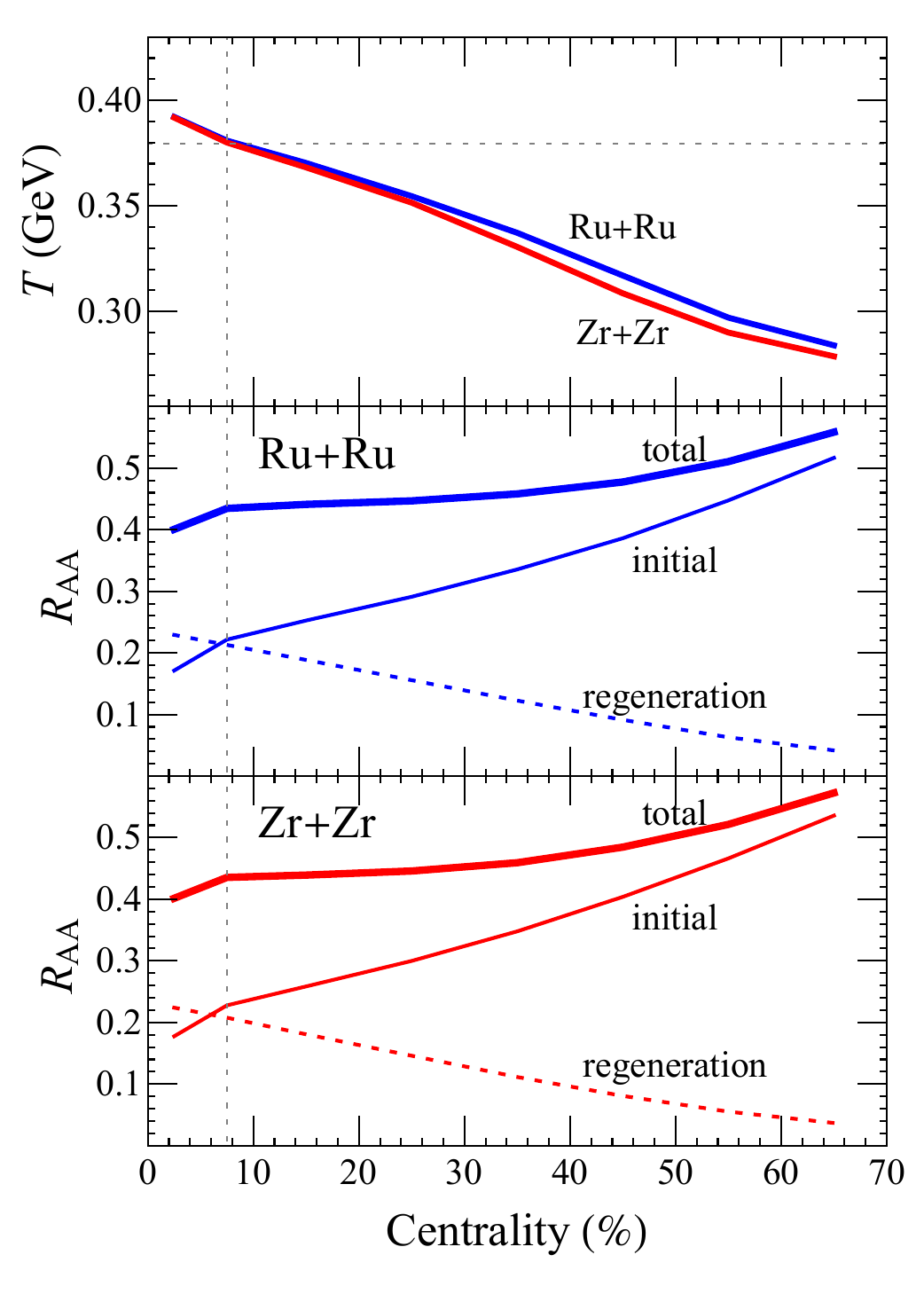}\qquad\qquad\qquad
\includegraphics[width=0.4\textwidth]{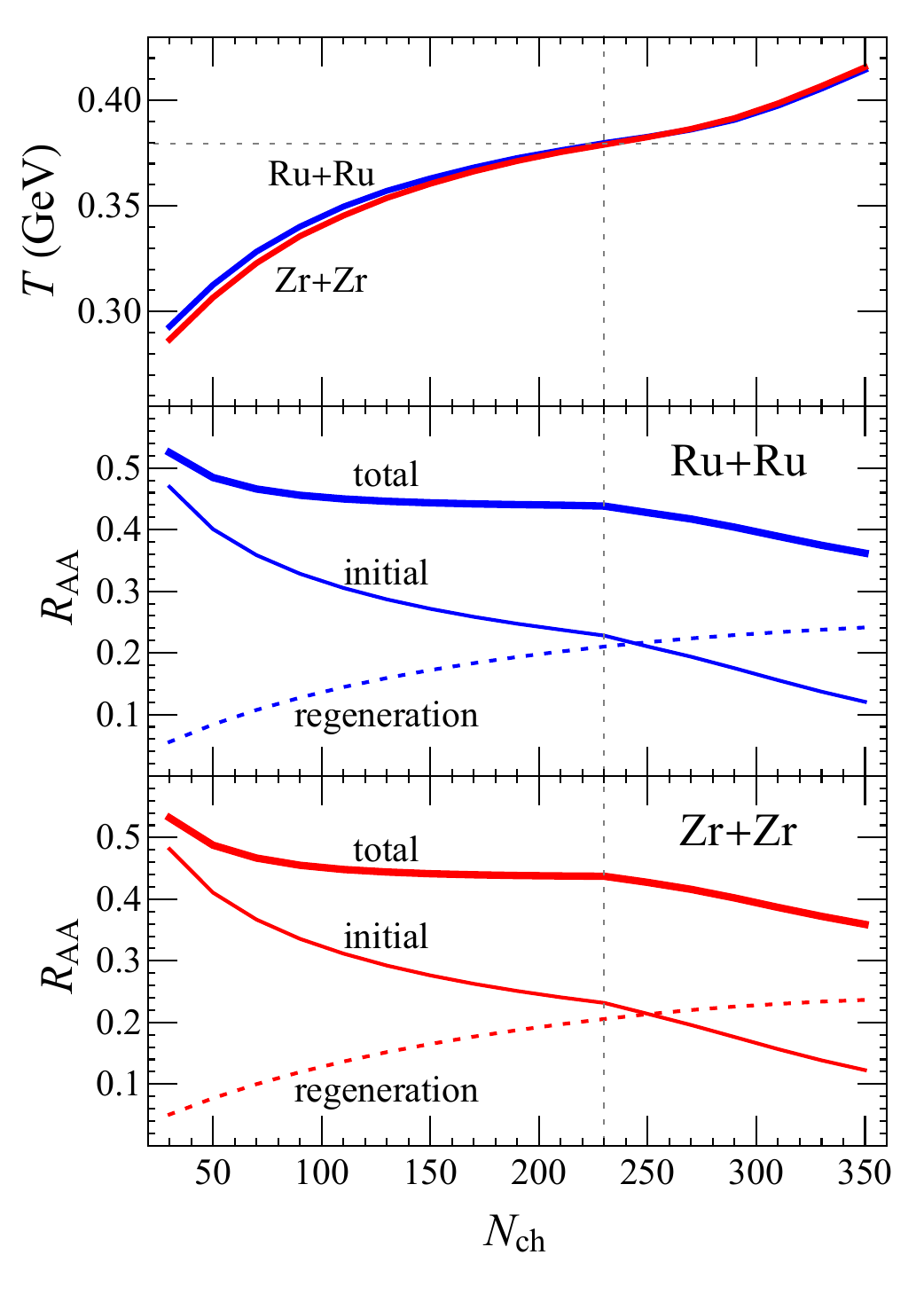}
\caption{(Left) The centrality dependence of the maximum temperature of the medium (upper), nuclear modification factor $R_{AA}$ of $J/\psi$ in Ru+Ru (middle) and Zr+Zr (lower) collisions at $\sqrt{s_\textrm{NN}} = 200~\textrm{GeV}$. Vertical dashed lines indicate the position of the cusp, whereas the horizontal dashed line label the dissociation temperature of $J/\psi$.
(Right) Same as Left but for equal-multiplicity bins. 
}
\label{fig2}
\end{figure*}

\vspace{5mm}
\emph{Charmonium transport.}---
We start by describing the charmonium dissociation and recombination in the QGP. Since charmonium states are heavy and color neutral, they are unlikely to be thermalized with the medium. Their phase space distribution, $f_\psi({\bm p},{\bm x},\tau)$ with $\psi \in \{J/\psi, \chi_c, \psi'\}$, is governed by a transport equation which includes both initial production and regeneration~\cite{Yan:2006ve,Liu:2009nb,Zhou:2014kka,Chen:2018kfo},
\begin{eqnarray}
\label{eq:transport}
&&\left[ \cosh(y-\eta)\partial_\tau + {\sinh(y-\eta)\over \tau}\partial_\eta+{\bm v}_T\cdot \nabla_T \right] f_\psi  \nonumber\\
&=& -\alpha\, f_\psi +\beta,
\end{eqnarray}
where $y=(1/2)\ln[(E+p_z)/( E-p_z)]$ is the momentum rapidity, ${\bm v}_T={\bm p}_T/E_T$ is the transverse velocity, $E_T=\sqrt{m_{\psi}^2+{\bm p}_T^2}$ is the transverse energy, and $\nabla_T\equiv(\partial_x, \partial_y)$ is the transverse gradient. The second(third) term on the left hand side arises from the free streaming of $\psi$ which leads to the leakage effect in the longitudinal(transverse) direction. The anomalous suppression and regeneration mechanisms in the QGP medium are respectively reflected in the loss term $\alpha$ and gain term $\beta$. Charmonia in hot QGP medium suffer Debye screening~\cite{Matsui:1986dk}. With increasing temperature, the interaction between a pair of heavy quarks is more screened, while the averaged size of a charmonium state increases. When the averaged size $\langle r\rangle$ diverges, the charmonium is dissociated.
We solve the two-body Schr\"odinger equation with the finite-temperature potential and define the dissociation temperature $T_d$ as the temperature when $\langle r\rangle\to \infty$. We find $T_d = (2.3, 1.2, 1.1)~T_c$ for $J/\psi$, $\chi_c$, and $\psi'$, respectively~\cite{Zhao:2022ggw}. When the local temperature is higher than the dissociation temperatures $T_d$, the charmonium state disappears, and the regeneration only happens when the temperature is lower than $T_d$. 
In addition to the Debye screening, charmonium states suffer dynamical dissociation in the QGP, such as gluon dissociation process, $g+\psi \to c + \bar c$~\cite{Peskin:1979va,Bhanot:1979vb}. The charmonium gluo-dissociation cross-section $\sigma_{g\psi}^{c\bar c}$ can be derived, via the operator-production-expansion (OPE) method, in vacuum~\cite{Peskin:1979va,Bhanot:1979vb} and extended to the finite temperature~\cite{Liu:2009nb,Zhou:2014kka,Chen:2018kfo}. Here we adopt the setup of~\cite{Liu:2009nb}. Taking only the gluo-dissociation as the loss term and its inverse process, $c+\bar c\to g+\psi$, as the gain term in this study, $\alpha$ and $\beta$ can be explicitly expressed as~\cite{Yan:2006ve}
\begin{eqnarray}
\label{eq:alphabeta}
\alpha({\bm p},{\bm x},\tau) &=& {1\over 2E_T}\int{d^3{\bm p}_g \over (2\pi)^3 2E_g} W_{g\psi}^{c\bar c}(T,s) f_g({\bm p}_g,{\bm x},\tau)\nonumber\\
&&\times\Theta(T({\bm x},\tau)-T_c),\nonumber\\
\beta({\bm p},{\bm x},\tau) &=& {1\over 2E_T}\int {d^3{\bm p}_g \over (2\pi)^3 2E_g}{d^3{\bm p}_c \over(2\pi)^3 2E_c}{d^3{\bm p}_{\bar c} \over(2\pi)^3 2E_{\bar c}}\nonumber\\
&&\times W_{c\bar c}^{g\psi}(T,s) f_c({\bm p}_c,{\bm x},\tau) f_{\bar c}({\bm p}_{\bar c},{\bm x},\tau)\nonumber\\
&&\times(2\pi)^4\delta^{(4)}(p+p_g-p_c-p_{\bar c})\nonumber\\
&&\times\Theta(T({\bm x},\tau)-T_c),
\end{eqnarray}
where $E_g$, $E_c$, and $E_{\bar c}$ are the gluon, charm quark and anti-charm quark energies, respectively, and ${\bm p}_g$, ${\bm p}_c$ and ${\bm p}_{\bar c}$ are their momenta. The Mandelstam variable $s$ is the $g\psi$ interaction energy. $W_{g\psi}^{c\bar c}$ is the dissociation probability while $W_{c\bar c}^{g\psi}$ the regeneration probability. Their related due to the detailed balance between the two processes, and the explicit expressions are given in~\cite{Polleri:2003kn}. $\Theta$ is the Heaviside step function to guarantee the calculation in the QGP phase above the critical temperature $T_c$. The latter is taken to be $T_c = 160~\text{MeV}$.

The gluon distribution $f_g$ follows the Bose-Einstein distribution,
whereas charm quarks with low transverse momenta are found to be thermalized according to experimental measurement of open-charm meson~\cite{STAR:2017kkh}, despite of their large mass. As a first-order approximation, we take a thermal distribution for charm/anti-charm momentum, $f_c({\bm p}_c,{\bm x},\tau) = {N_c \rho_c({\bm x},\tau) \over e^{p_c^\mu u_\mu/T}+1}$, where $N_c$ is the normalization factor to ensure $\int {N_c \over e^{p_c^\mu u_\mu/T}+1} {d^3{\bm p}_c \over(2\pi)^3 2E_c} = 1$. 
Considering the small transverse velocity and the similarity of the isobar system,  we neglect the density change due to transverse expansion, and the density in coordinate space is governed by the nuclear geometry of the colliding system~\cite{Liu:2009nb},
\begin{eqnarray}
\label{charm}
\rho_c({\bm x},\tau)&=&{\rho_\mathrm{coll}({\bm x})}
{\cosh \eta \over \tau}{d\sigma_{pp}^{c\bar c}\over dy},
\end{eqnarray}
where $d\sigma^{c\bar{c}}_{pp}/dy$ is charm quark production cross section per unit rapidity in p+p collisions. We take $d\sigma^{c\bar{c}}_{pp}/dy = 0.162 ~\mathrm{mb}$, which corresponds to $\sqrt{s_\mathrm{NN}}=200~\text{GeV}$~\cite{Zhao:2022ggw}. 
The binary collision density $\rho_\mathrm{coll}({\bm x})$ is given by the initial condition of the bulk background. In this work, we adopt the boost invariant, event-by-event fluctuating initial conditions generated by the Monte Carlo Glauber model (MCGlauber)~\cite{Miller:2007ri}. We generate millions of initial profiles and bin them into different centrality or multiplicity classes. Events within the same class are then averaged after being overlapped according to the center of mass and aligned according to the second-order participant plane. With such event-averaged initial conditions, we solve the medium evolution taking the data validated \textsc{music} hydrodynamic simulation package~\cite{Gale:2013da,Schenke:2010rr,Schenke:2010nt,McDonald:2016vlt}, with taking the \textsc{s95p} Equation of State~\cite{Huovinen:2009yb}, a constant shear viscosity $\eta/s$ = 0.08, and vanishing bulk viscosity~\cite{Policastro:2001yc,Kovtun:2004de,Bernhard:2016tnd}. The hydrodynamic evolution provides the space-time profile of temperature $(T)$ and flow velocity $(u^\mu)$ of the bulk background in the calculation of charmonium transport and gluon and charm quark distribution.
The centrality and multiplicity dependence of the maximum temperature of the medium produced in Ru+Ru and Zr+Zr collisions are shown in the upper panels of Fig.~\ref{fig2}. One can find them to be close to identical in central collisions but visibly different in peripheral collisions.

\begin{figure*}[!htb]
\includegraphics[width=0.4\textwidth]{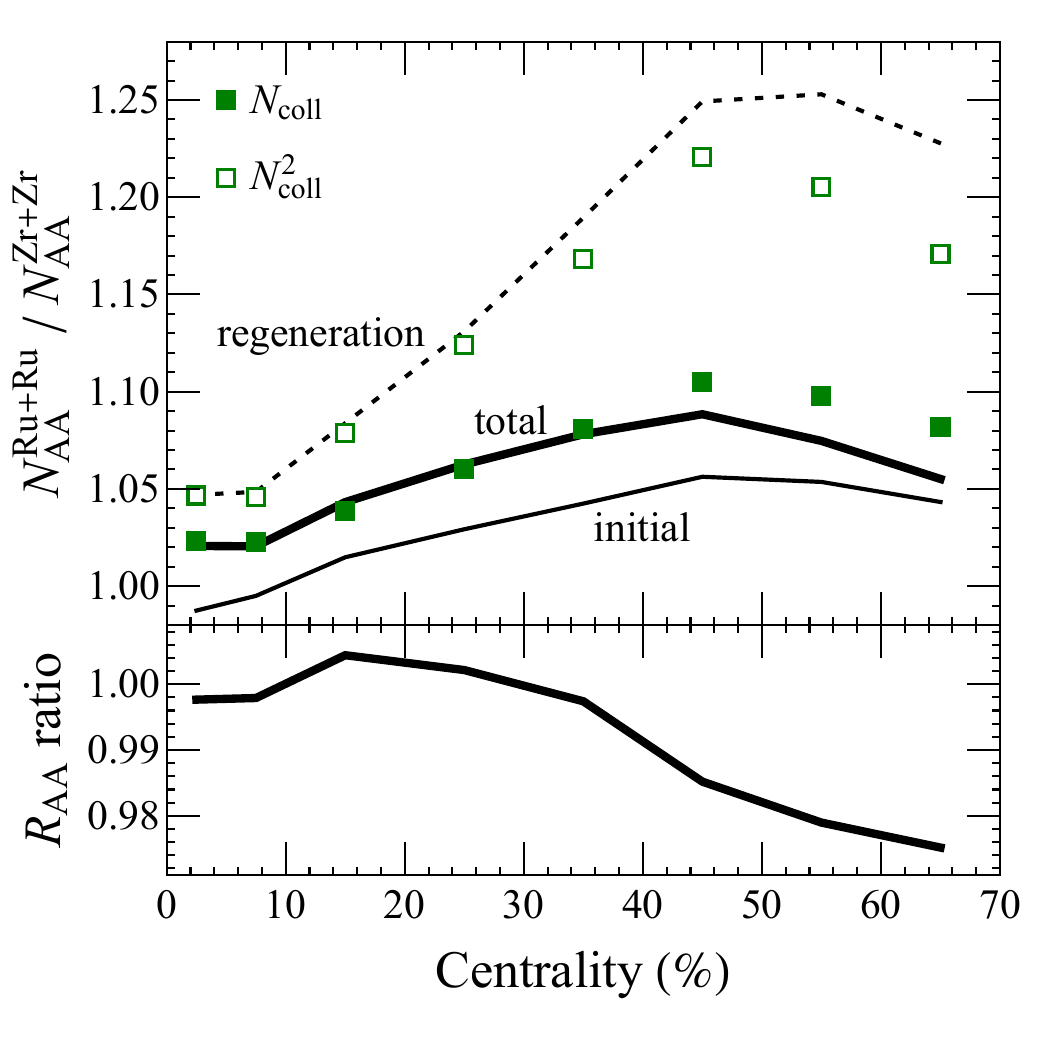}\qquad\qquad\qquad
\includegraphics[width=0.4\textwidth]{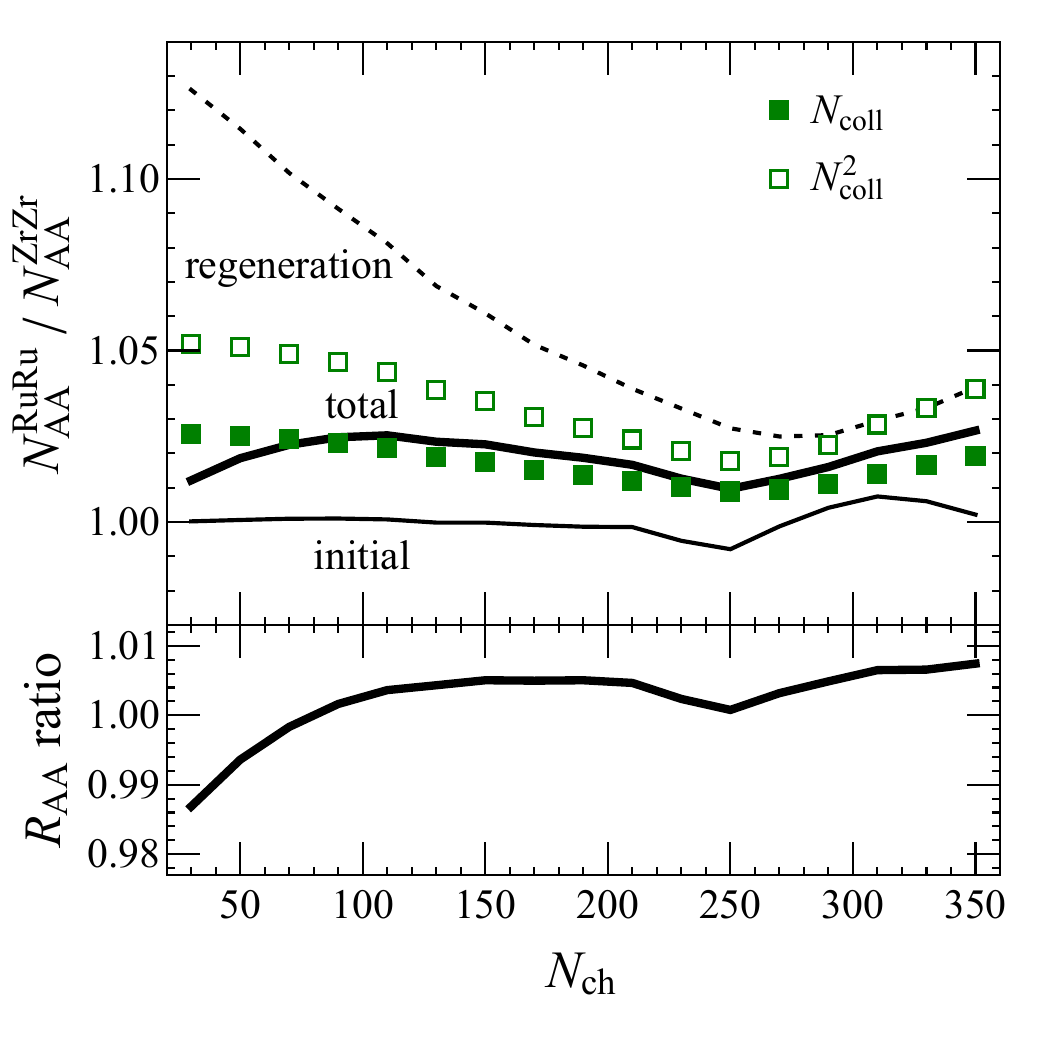}
\caption{The centrality  (left) and multiplicity (right) dependence of the yield ratio ($N_{\text{AA}}^{\text{RuRu}}/N_{\text{AA}}^{\text{ZrZr}}$) and double ratio ($R_{\text{AA}}^{\text{RuRu}}/R_{\text{AA}}^{\text{ZrZr}}$) of prompt $J/\psi$ produced in Ru+Ru and Zr+Zr collisions at $\sqrt{s_\textrm{NN}} = 200~\textrm{GeV}$. Dotted lines represent the regeneration contribution. While thin and thick solids lines are the initial and total results, respectively. The filled dots are the ratio of number of collisions $N_{\text{coll}}$ between Ru+Ru and  Zr+Zr, and open dots are its square.}
\label{fig3}
\end{figure*}
Besides the hot nuclear matter effect which affects charmonium motion through the above discussed anomalous suppression and regeneration in the QGP, there is also the cold nuclear matter effect which changes the initial condition of the transport equation~\eqref{eq:transport}. The cold nuclear matter effect includes mainly the nuclear absorption~\cite{Gerschel:1988wn}, Cronin effect~\cite{Cronin:1974zm}, and nuclear shadowing~\cite{Mueller:1985wy}. Considering a finite formation time of charmonia and small nuclear size of Ru and Zr in high energy collisions, the nuclear absorption is neglected in this study. 
Such an approximation can be further justified by the fact that the absorption effect shall be similar for both nuclei.
The initial charmonium distribution in heavy-ion collisions is constructed by a superposition of the charmonium distribution in p+p collisions with considering the cold nuclear matter effects,
\begin{eqnarray}
\begin{split}
f_\psi({\bm p}, {\bm x}, \tau_0) 
=\;& {(2\pi)^3 \over E \tau_0}\rho_{\text{coll}}({\bm x})
f^\mathrm{Cronin}_\psi({\bm p},{\bm x})\\
&\times \mathcal{R}_g^P({\bm p},{\bm x})\mathcal{R}_g^T({\bm p},{\bm x}),
\end{split}
\label{eq:initial}
\end{eqnarray}
where the differential cross section~\cite{Zha:2015eca,Zhao:2022ggw},
\begin{eqnarray}
\label{initialpp}
f^\mathrm{Cronin}_\psi
= {a\over 2\pi\langle p_T^2\rangle_A}\left(1+b^2{p_T^2\over \langle p_T^2\rangle_A}\right)^{-n} {d\sigma^{\psi}_{pp}\over dy},
\end{eqnarray}
encodes the Cronin effect that enhance the transverse momentum compared to p+p collisions
\begin{eqnarray}
\langle p_T^2\rangle_A \approx \langle p_T^2\rangle_p+a_{gN}\,\rho_0^{-1}\sqrt{\rho_{\text{coll}}({\bm x})/\sigma_\textrm{NN}^{inel}}.
\end{eqnarray}
The parameters $a=2b^2(n-1)$, $b=\Gamma(\frac{3}{2})\Gamma(n-\frac{3}{2})/\Gamma(n-1)$, $n=3.93$, and $\langle p_T^2\rangle_p = 3.05~\textrm{GeV}^2$ are fitted to match the experimental data for p+p collisions~\cite{Zha:2015eca,Zhao:2022ggw,STAR:2018smh}.
The Cronin parameter, which describes the averaged charmonium transverse momentum square obtained from the gluon scattering with a unit of length of nucleons, is set as $a_{gN}=0.1~\textrm{GeV}^2/\fm$ ~\cite{Zhao:2010nk,Zhao:2022ggw} in order to fit experimental data in p-A and low-energy A-A collisions~\cite{NA50:2003tdy,PHENIX:2007tnc}. The mean trajectory length of the two gluons in the two nuclei before the $c\bar c$ formation has been estimated as $\rho_0^{-1}\sqrt{\rho_{\text{coll}}({\bm x})/\sigma_\textrm{NN}^{inel}}$, where $\rho_0$ is the maximum nuclear density in Eq.~\eqref{eq.ws}. The direct production cross sections 
$d\sigma^{\{J/\psi,\chi_c,\psi'\}}_{pp}/dy = \{0.6, 0.3, 0.1\}\times d\sigma^{J/\psi,\mathrm{prompt}}_{pp}/dy$~\cite{Andronic:2015wma}, where 
$d\sigma^{J/\psi,\mathrm{prompt}}_{pp}/dy=716.7~\mathrm{nb}$ is  prompt production cross section~\cite{STAR:2018smh}.

$\mathcal{R}_g^P$ and $\mathcal{R}_g^T$ in Eq.~(\ref{eq:initial}) are spatial-dependent shadowing factors for gluon in projectile and target. Assuming that the inhomogeneous shadowing is proportional to the parton path length through the nucleus, the spatial-dependent shadowing factor can be estimated as,
\begin{equation}
\mathcal{R}_g^{P,T}({\bm p},{\bm x}) \approx 1+
{A[R_g(x_g^{P,T}, Q^2)-1] \sqrt{\rho_{\text{coll}}({\bm x})} \over
N_\text{coll}^\text{cen} /\sqrt{\sigma_\textrm{NN}^{inel}}},
\end{equation}
where the shadowing modification factor $R_g$ can be simulated by the EPS09 package~\cite{Helenius:2012wd}.
$x_{g}^{P,T}=(\sqrt{m_\psi^2+p_T^2}/\sqrt{s_\textrm{NN}})e^{\pm y}$ is gluon longitudinal momentum fraction with $y$ being the charmonium rapidity, and the $+$($-$) sign is chose for projectile(target). The momentum transfer squared can be taken as $Q^2 = m_{\psi}^2+p_T^2$. For both Ru and Zr, the nuclear mass number is $A=96$. $N_\text{coll}^\text{cen}$ is the number of binary collision for the most central collisions.

\vspace{5mm}
\emph{Results.}---
Experimental measurement of prompt $J/\psi$ mesons include feed-down contributions from the excited charmonium states, i.e., $\chi_c$ and $\psi'$. 
With initial condition given in Eq.~\eqref{eq:initial}, we solve the transport equations~\eqref{eq:transport} for $J/\psi$, $\psi'$ and $\chi_c$ on top of event-averaged hydrodynamic background corresponding to different collision systems and different centrality or multiplicity bins. Then we decay $\chi_c$ and $\psi'$ into $J/\psi$, with branch ratios respectively being $22\%$ and $61\%$~\cite{ParticleDataGroup:2020ssz}, and obtain the properties of prompt $J/\psi$ in the final state.
It shall be noted that the non-prompt contribution from $B$-decays can be neglected at $200~\textrm{GeV}$ Ru+Ru and Zr+Zr collisions, due to the small production cross-section of the latter at low beam energy~\cite{Andronic:2015wma}.
Therefore, inclusive production rate of $J/\psi$ is dominated by the prompt one.

We first study the nuclear modification factor of $J/\psi$ in both Ru+Ru and Zr+Zr collisions, which is defined as the ratio of production rate in AA to that in p+p collisions, scaled by the inverse of number of binary collisions,
\begin{align}
    R_\text{AA} = \frac{N_\text{AA}}{N_{pp}\, N_\text{coll}}\,.
\end{align}
In Fig.~\ref{fig2}, we present both the centrality and multiplicity dependence of prompt $J/\psi$ $R_{AA}$ in Ru+Ru and Zr+Zr collisions at $\sqrt{s_\textrm{NN}} = 200~\textrm{GeV}$, with contributions from initial production and regeneration separated. 
We observe a clear cusp located the $5-10\%$ centrality bin or the $\sim 230$ multiplicity bin. 
For collisions more central than the cusp, the highest temperature is higher than the dissociation temperature of $J/\psi$, $T_d = 2.3~T_c$, which is indicated by the horizontal dashed lines. $J/\psi$ mesons initially produced at the $T>T_d$ region suffer a strong suppression.
Meanwhile, one can find the regeneration contribution plans an important role in final total yield, especially at central collisions even in small systems like Ru+Ru and Zr+Zr collisions. 

Then we move on to compare the $J/\psi$ production in the isobar systems by taking the ratio between observables in Ru+Ru and those in Zr+Zr collisions. In the upper panels of Fig.~\ref{fig3}, we show the ratio of total yield as well as those for initial and regeneration productions. In the hypothetical case of identical background in the isobar systems, one would expect the ratio of initial production (thin solid lines) to be the same as that of $N_\text{coll}$ (filled square), whereas the ratio of regeneration production (thin dash lines) to be identical to that of $N_\text{coll}^2$ (open square)\footnote{It shall be worth noting that the difference due to cold nuclear effect is negligible}. We observe that the former is below the $N_\text{coll}$ ratio, because higher temperature and larger volume in Ru+Ru collisions. Meanwhile, the ratio of regeneration production agree well with the $N_\text{coll}^2$ ratio at central collisions, and goes above such a expectation for peripheral collisions owning to the longer evolution time in Ru+Ru collisions, and therefore more $J/\psi$ particles are produced via regeneration. As for the double ratio, by definition, it equals to the yield ratio times the inverse ratio of the $N_\text{coll}$. So, the double ratio is smaller than 1 and decreases monotonously at peripheral collisions.
In comparing the left and right panels, one can clearly see that all ratios are closer to unity when we compare the isobar system within the same multiplicity bin than the comparison for the same centrality bin.

\begin{figure}[!htb]
\includegraphics[width=0.4\textwidth]{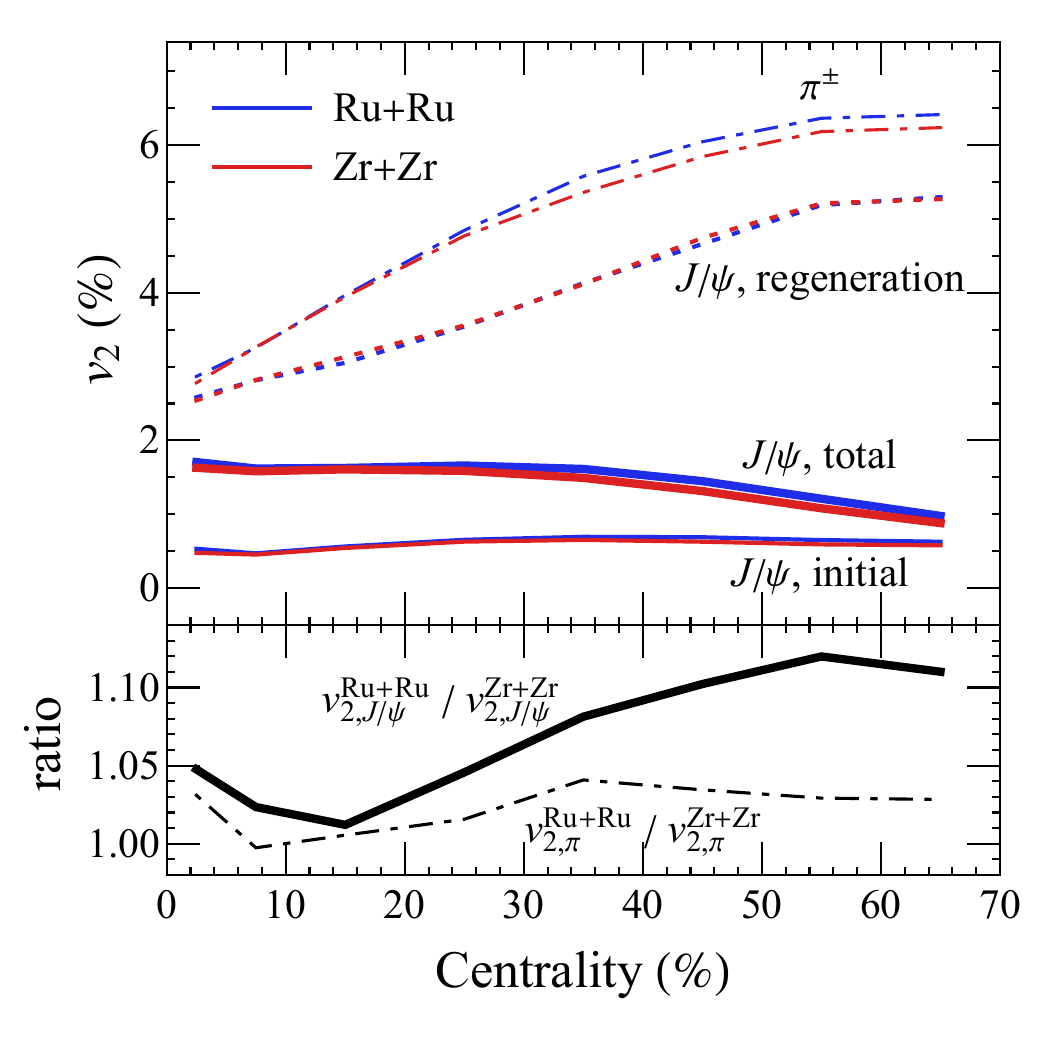}
\caption{(Upper panel) The centrality dependence of the elliptic flow $v_2$ of light hadrons $\pi$ and $J/\psi$ produced in Ru+Ru and Zr+Zr collisions at $\sqrt{s_\textrm{NN}} = 200~\textrm{GeV}$. (Lower panel) The ratio of $v_2$ in Ru+Ru and Zr+Zr collisions for $J/\psi$ and $\pi^{\pm}$.}
\label{fig4}
\end{figure}

The nucleon distribution in the isobar nuclei influences the anisotropy of the medium created in heavy-ion collisions. The latter can be measured by the momentum anisotropy of final state hadrons, which is defined as the Fourier coefficient of their azimuthal distribution and referred to as $v_n$~\cite{Ollitrault:1992bk}. 
The second flow coefficient, called elliptic flow $v_2$, is mostly a collective flow response to the ellipsoidal shape of the overlap region in non-central collisions. 
The difference in the light flavor $v_2$ has been found to be sensitive to the initial anisotropy contributed by the deformation~\cite{Zhang:2021kxj,Jia:2021tzt}, and it would be interesting to measure $v_2^{J/\psi}$ which probes the medium anisotropy in a different way. Results are shown in Fig.~\ref{fig4}.
The $v_2$ of $J/\psi$ is influenced by two origins. One is regeneration in which the anisotropy is similar to that of $\pi$ in both the centrality trend and order of magnitude. The other is the initial production which is isotropic initially and pick up some small amount of anisotropy via path-dependent dissociation in the anisotropic medium background. In central collisions, the anisotropy of the background is small, whereas in peripheral collisions the temperature is small and the dissociation effect is weak. Both these factors make $v_2$ of initially produced $J/\psi$ flat and small for all centrality ranges. 
As is shown in Fig.~\ref{fig2}, going from the most central to peripheral collisions, the regeneration fraction decreases and henceforth the overall $v_2$ of $J/\psi$ decreases, which is opposite to the trend of light hadrons.
A small tilde (enhancement) in $v_2$ for the most central collisions is also observed, because the existence of $T>T_d$ region induces extra anisotropy for initially produced $J/\psi$. With the Ru-to-Zr ratio shown in the lower panel, we expect $\sim5-10\%$ difference in the elliptic flow of $J/\psi$. It shall be worth noting that such a ratio is greater than that of $v_{2,\pi}$. With details shown in the Appendix, the difference of $v_{2,J/\psi}$ in the isobar system is contributed dominantly by the fact that there is a greater portion of $J/\psi$ that is created via regeneration production in Ru+Ru collisions than in Zr+Zr. Thus, a precise measurement of $v_{2,J/\psi}$ difference serves as an accurate probe of the regeneration fraction.

\vspace{5mm}
\emph{Summary.}---
Via a relativistic transport approach, we investigate the yield and elliptic flow of $J/\psi$ in isobar collisions. The evolution of the hot medium produced in Ru+Ru and Zr+Zr collisions are described by relativistic viscous hydrodynamics with event averaged initial conditions. Aiming to detect the difference in nucleon distributions between Ru and Zr, we compare the observables in the isobar system according to the same centrality bin and the same multiplicity bin. Due to the charm produced via initial binary collisions, the yield ratio can clearly show the difference between Ru and Zr. 
The elliptic flow $v_2$ of $J/\psi$ in isobar collisions is qualitatively different from the light hadrons, and the ratio between Ru+Ru and Zr+Zr collisions is around $5-10\%$ in non-central collisions. These heavy flavor particles serve as an independent probe to characterize the number of binary collisions in the initial state and therefore the nucleon distribution in the isobar nuclei.

The regeneration production of charmonium is proportional to the square of the production cross-section of charm quark, with the latter being proportional to the number of binary collision and also increasing with collision energy. At LHC energy charmonium states are produced dominantly via regeneration and we expect the difference in nucleon distribution to be better reflected there.

In addition, the deformation of the nucleus is better reflected in ultra-central collisions, in which one may select events with preference on the so called body-body collisions. In such events, the deformation axes of the projectile and target are parallel to each other, and both of them are perpendicular to the beam direction. Therefore, the deformation of the nucleon distribution is translated into the eccentricity of the initial condition, and henceforth the final state anisotropy. Comparison of both light and heavy flavor observables will be reported in our future publication.

\vspace{5mm}
{\bf Acknowledgement:} We thank Haojie Xu and Pengfei Zhuang for helpful discussion. J.Z. is supported by the NSFC Grant Nos. 12175165 and 12047535, and the European Union's Horizon 2020 research and innovation program under grant agreement No 824093 (STRONG-2020). S.S. acknowledges support by the U.S. Department of Energy, Office of
Science, Office of Nuclear Physics, Grants Nos. DE-FG88ER41450 and DE-SC0012704.

\begin{appendix}
\begin{widetext}
\section{Source of difference in elliptic flow}
\begin{figure}[!hbt]
    \centering
    \includegraphics[width=0.4\textwidth]{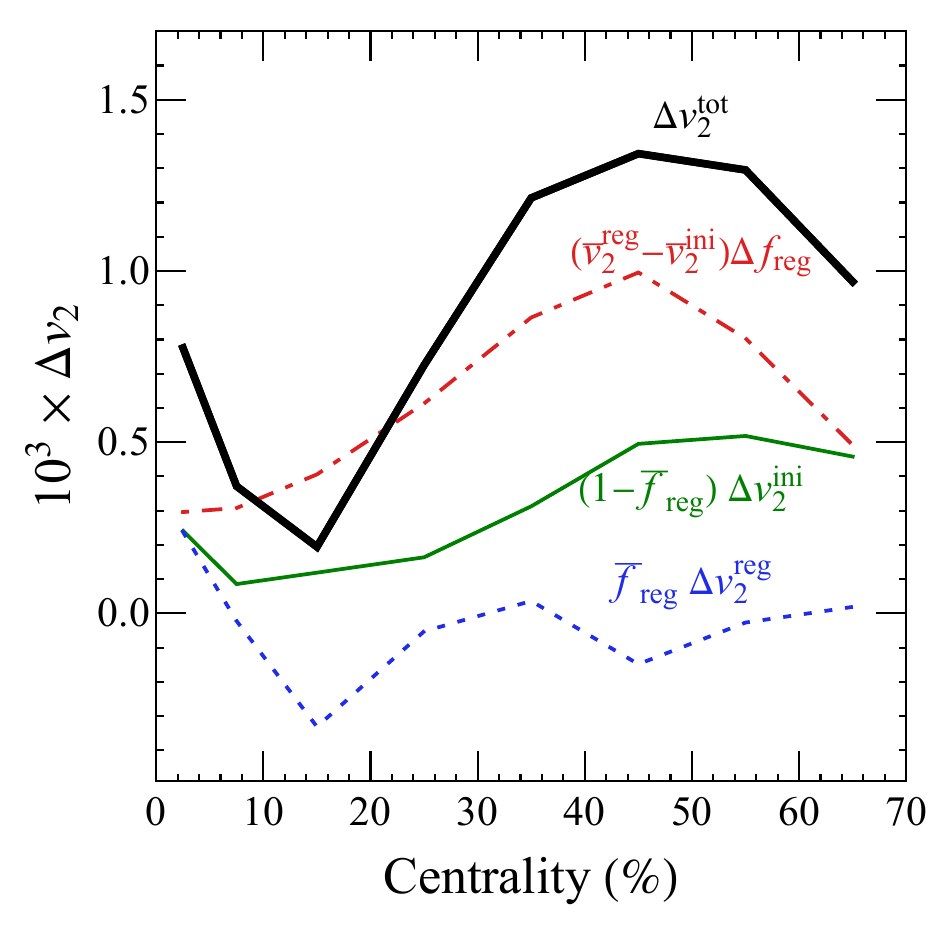}\qquad
    \includegraphics[width=0.4\textwidth]{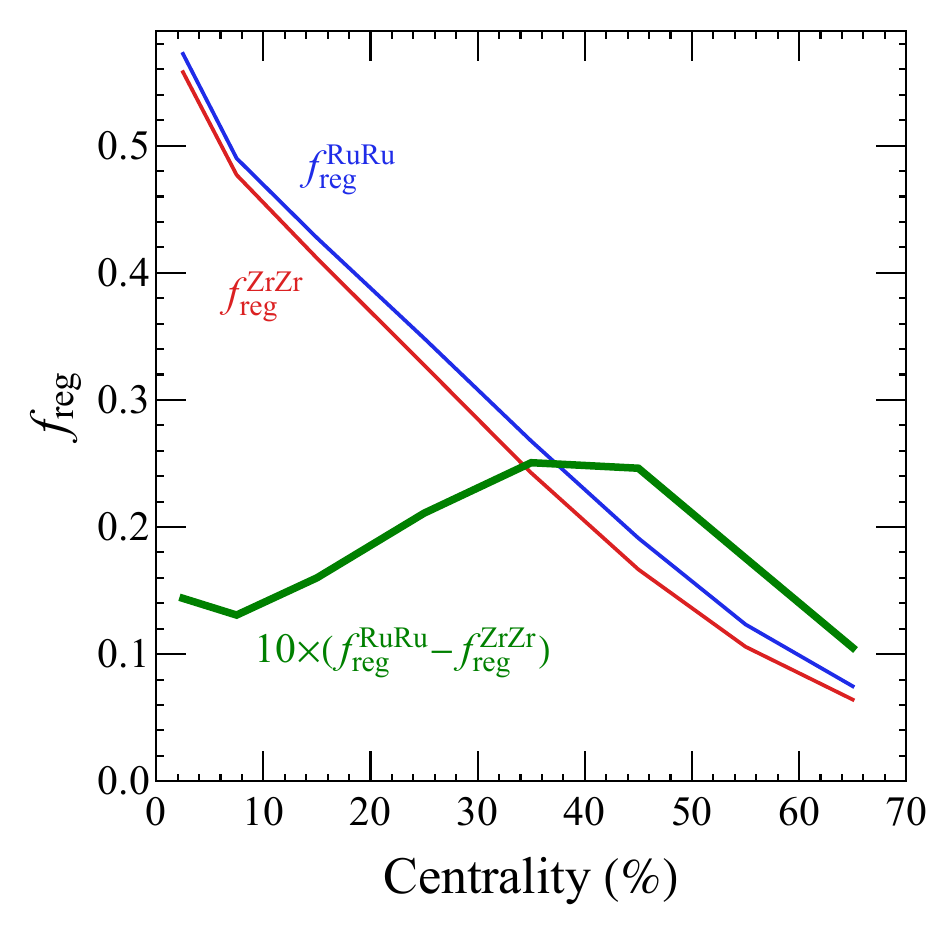}
    \caption{(Left) Sources of $v_{2,J/\psi}$ difference. (Right) Fraction of regeneration production in the final state $J/\psi$ in isobar collisions.}
    \label{fig.v2_diff}
\end{figure}
In this appendix, we discuss the source of difference in $v_{2,J/\psi}$ between Ru+Ru and Zr+Zr collisions. We separately compare the $v_2$ of initial production and regeneration productions. Denoting $f_\mathrm{reg}$ as the fraction of regeneration production in all the final state $J/\psi$ particles, one can perform the separation as $v_{2,\mathrm{tot}} = (1-f_\mathrm{reg}) v_{2,\mathrm{ini}} + f_\mathrm{reg} v_{2,\mathrm{reg}}$. Then, one can show that
\begin{align}
\begin{split}
&\Delta v_{2}^\mathrm{tot} 
\\=\;&
(1-f_\mathrm{reg}^\mathrm{Ru}) v_{2,\mathrm{ini}}^\mathrm{Ru}
+ f_\mathrm{reg}^\mathrm{Ru} v_{2,\mathrm{reg}}^\mathrm{Ru}
- (1-f_\mathrm{reg}^\mathrm{Zr}) v_{2,\mathrm{ini}}^\mathrm{Zr}
- f_\mathrm{reg}^\mathrm{Zr} v_{2,\mathrm{reg}}^\mathrm{Zr}\\
=\;&
\Big(1-\frac{f_\mathrm{reg}^\mathrm{Ru}+f_\mathrm{reg}^\mathrm{Zr}}{2}\Big) (v_{2,\mathrm{ini}}^\mathrm{Ru} - v_{2,\mathrm{ini}}^\mathrm{Zr})
- (f_\mathrm{reg}^\mathrm{Ru}-f_\mathrm{reg}^\mathrm{Zr}) \frac{v_{2,\mathrm{ini}}^\mathrm{Ru} + v_{2,\mathrm{ini}}^\mathrm{Zr}}{2}
+ \frac{f_\mathrm{reg}^\mathrm{Ru}+f_\mathrm{reg}^\mathrm{Zr}}{2} (v_{2,\mathrm{reg}}^\mathrm{Ru} - v_{2,\mathrm{reg}}^\mathrm{Zr})
+ (f_\mathrm{reg}^\mathrm{Ru}-f_\mathrm{reg}^\mathrm{Zr}) \frac{v_{2,\mathrm{reg}}^\mathrm{Ru} + v_{2,\mathrm{reg}}^\mathrm{Zr}}{2}
\\
=\;&
\Big(1-\frac{f_\mathrm{reg}^\mathrm{Ru}+f_\mathrm{reg}^\mathrm{Zr}}{2}\Big) (v_{2,\mathrm{ini}}^\mathrm{Ru} - v_{2,\mathrm{ini}}^\mathrm{Zr})
+ \frac{f_\mathrm{reg}^\mathrm{Ru}+f_\mathrm{reg}^\mathrm{Zr}}{2} (v_{2,\mathrm{reg}}^\mathrm{Ru} - v_{2,\mathrm{reg}}^\mathrm{Zr})
+ (f_\mathrm{reg}^\mathrm{Ru}-f_\mathrm{reg}^\mathrm{Zr}) \Big(\frac{v_{2,\mathrm{reg}}^\mathrm{Ru} + v_{2,\mathrm{reg}}^\mathrm{Zr}}{2} - \frac{v_{2,\mathrm{ini}}^\mathrm{Ru} + v_{2,\mathrm{ini}}^\mathrm{Zr}}{2} \Big)
\\\equiv\;&
(1-\bar{f}_\mathrm{reg}) \Delta v_{2,\mathrm{ini}}
+ \bar{f}_\mathrm{reg} \Delta v_{2,\mathrm{reg}}
+  (\bar{v}_{2,\mathrm{reg}} - \bar{v}_{2,\mathrm{ini}} )\Delta f_\mathrm{reg}\,.
\end{split}
\end{align}
In the left panel of Fig.~\ref{fig.v2_diff}, we plot these three sources separately, together with the full value of $\Delta v_{2}^\mathrm{tot}$. It turns out that the last term, which is caused by the fact that there is a greater portion of $J/\psi$ that is created via regeneration production in Ru+Ru collisions than in Zr+Zr. As noted in the main text that ${v}_{2,\mathrm{reg}}$ is much greater than ${v}_{2,\mathrm{ini}}$, the difference in $f_\mathrm{reg}$ makes the greatest contribution to the difference in elliptic flow.

\end{widetext}
\end{appendix}

\bibliographystyle{apsrev4-1.bst}
\bibliography{Ref}

\end{document}